\documentclass[aps,prb,twocolumn,showpacs,floatfix]{revtex4}

\usepackage{graphicx}

\begin{document}

\title{The electronic structure of liquid water 
       within density functional theory}
\author{David Prendergast, Jeffrey C. Grossman, and Giulia Galli} 
\affiliation{Lawrence Livermore National Laboratory, L-415, P.O.  Box 808, 
             Livermore, CA 94551.}

\date{\today}


\begin{abstract}
In the last decade, computational studies of liquid water have mostly
concentrated on ground state properties. However recent spectroscopic 
measurements have been used to infer the structure of water, and the 
interpretation of optical and x-ray spectra requires accurate theoretical 
models of excited electronic states, not only of the ground state. 
To this end, we investigate the electronic properties of water
at ambient conditions using \emph{ab initio} density functional theory
within the generalized gradient approximation (DFT/GGA),
focussing on the unoccupied subspace of Kohn-Sham eigenstates.
We generate long (250 ps) classical trajectories for large supercells,
up to 256 molecules, from which uncorrelated configurations
of water molecules are extracted for use in DFT/GGA calculations of
the electronic structure.
We find that the density of occupied states of
this molecular liquid is well described with 32 molecule supercells
using a single k-point (${\bf k}={\bf 0}$)
to approximate integration over the first Brillouin zone.
However, the description of the density of \emph{unoccupied}
states (u-EDOS) is sensitive to finite size effects.
Small, 32 molecule supercell calculations, using the $\Gamma$-point 
approximation, yield a spuriously isolated state above the Fermi level.
Nevertheless, the more accurate u-EDOS of large, 256 molecule supercells
may be reproduced using smaller supercells and increased k-point sampling.
This indicates that the electronic structure of molecular liquids like
water is relatively insensitive to the long-range disorder in the molecular
structure.
These results have important implications for efficiently increasing 
the accuracy of spectral calculations for water and other molecular liquids.
\end{abstract}


\maketitle

\section{Introduction}
\label{Sec.intro}

Our understanding of the structure of water in its many phases is
fundamental to research in fields as diverse as biochemistry, cellular biology, atmospheric chemistry,
and planetary physics.
The standard experimental approaches to
the analysis of crystal structures -- x-ray~\cite{Hura_Russo_2003_pccp}
and neutron~\cite{Soper_2000_chemphys} diffraction -- can
provide detailed information on ordered phases of ice, but only indirect, 
and often limited, information on amorphous ice and liquid water.
Nonetheless, the results of these experiments have permitted major advances in the understanding of the properties of water in the last 30 years; in addition they made possible the development of simple classical
potentials to describe water in the condensed phase, thus enabling
molecular dynamics simulations to probe its dynamical properties.

Recently, more sophisticated {\it ab initio} electronic structure approaches 
have become available to perform simulations of 
water.~\cite{Car_Parrinello_1985_prl}
{\it Ab initio} molecular 
dynamics based on density functional 
theory~\cite{Hohenberg_Kohn_1964_pr,Kohn_Sham_1965_pr} 
(DFT) has been extensively used to 
study water and solvation processes. It is not yet fully understood how 
accurate DFT is~\cite{Asthagiri_Pratt_2003_pre,Grossman_Schwegler_2004_jcp}
-- and, in particular, how accurate the various 
gradient corrected
functionals~\cite{Langreth_Mehl_1983_prb,Becke_1988_pra,
                  Perdew_Chevary_1992_prb} 
are -- in describing the structural and diffusive properties of liquid water. 
Furthermore, the quantitative influence of the inclusion of proton 
quantum effects in 
{\it ab initio} simulations remains to be 
established.~\cite{Schwegler_Grossman_2004_jcp}  
However, recent results 
reported in the literature have shown that qualitative features of 
hydrogen bonding in liquid water are correctly accounted for by {\it ab initio} 
molecular dynamics (MD) based on DFT.~\cite{Grossman_Schwegler_2004_jcp,
Schwegler_Grossman_2004_jcp,VandeVondele_Mohamed_2005_jcp}

The majority of {\it ab initio} studies, to date, have concentrated on 
structural 
properties of liquid water, and, only recently, 
have first-principles calculations of 
absorption spectra been carried out.~\cite{Cavalleri_Ogasawara_2002_cpl,
Myneni_Luo_2002_jpcm,Wernet_Nordlund_2004_science,
Hetenyi_DeAngelis_2004_jcp,Nordlund_Ogasawara_2004_cpl,
Cavalleri_Odelius_2004_jcp,Cai_Mao_2005_prl} 
These computations were motivated by new spectroscopic results obtained using 
x-ray Raman spectroscopy~\cite{Bowron_Krisch_2000_prb,Bergmann_Wernet_2002_prb,
Cai_Mao_2005_prl} 
and x-ray absorption spectroscopy~\cite{Myneni_Luo_2002_jpcm,
Parent_Laffon_2002_jcp,Wernet_Nordlund_2004_science,
Nordlund_Ogasawara_2004_cpl} 
for water and ice; and
x-ray absorption spectroscopy for liquid water jets.~\cite{Smith_Cappa_2004_science,
Wilson_Rude_2004_rsi}  
The experimental spectra -- providing direct information 
on the electronic transitions from atomic core levels to excited states --
have been used to infer information on the structural properties of the fluid.
In particular, the number of hydrogen bonds and the details of the 
hydrogen bonded network in the fluid have been inferred from experiment
and compared to those of ice. 
A standard procedure has been to use DFT in the gradient corrected 
approximation to compute absorption spectra of selected snapshots representing 
the liquid -- within some approximation for the description of the core-hole 
interaction. Those snapshots producing spectra in agreement with experiment 
have then been considered to be representative of the ``correct'' water 
structure, as probed experimentally. While this may well be a viable and 
straightforward approach in the absence of major approximations in the theory 
used to compute spectra, it becomes a much more complex interpretative tool 
in the presence of approximations in the theory, in particular approximations
regarding the core-hole interaction with the excited electron.

In order to provide a clear interpretation of measured and computed spectra, 
an important prerequisite is to fully understand the electronic structure of 
the fluid, as described within DFT. The purpose of this article is to provide 
a detailed  description of the
electronic structure of liquid water using
Kohn-Sham density functional theory, and in particular to understand in 
detail the
unoccupied subspace of Kohn-Sham eigenstates -- especially the
conduction band minimum, which will have a great influence in determining the 
onset of computed absorption spectra.  

We shall show that the electronic density of states (EDOS) of liquid water,
near the conduction band minimum, 
is particularly sensitive to finite size effects.
We find that for small, 32 molecule supercell calculations of the DFT
electronic structure of liquid water, approximating Brillouin zone 
integration by sampling the $\Gamma$-point only, leads to an EDOS
where the lowest unoccupied state is separated from the rest of the
conduction band by $\sim 1.5$~eV. This is consistent with previous
DFT calculations of the electronic structure of water.

In order to understand this peculiar feature of the EDOS of water, we
first examine the isolated dimer and the ordered
hexagonal phase of ice I$h$. The Kohn-Sham eigenstate energies of
the dimer indicate a separation in energy of the lowest unoccupied molecular
orbital (LUMO) from the rest of the unoccupied states. 
This is apparently consistent with the separation of such states 
in liquid water.
However, water possesses a hydrogen bonded network more similar to ice
than to the isolated dimer.
An examination of the band structure of the ice I$h$ crystal indicates
large dispersion in the conduction band states. In particular, there is
a large separation ($\sim 3$~eV) between the two lowest conduction bands
at the $\Gamma$-point. 

Inspired by these observations, we examine the convergence of the EDOS
of liquid water with respect to k-point sampling.
We compute the EDOS using several uncorrelated configurations of 
water molecules generated using molecular dynamics at ambient conditions
with a classical potential.
Examining the EDOS of several liquid water configurations, 
we find that the occupied portion is converged using just one k-point.
This implies that the results of previous DFT studies of water
in the electronic ground state would be unaffected by increased k-point 
sampling. However, for the unoccupied portion of the EDOS,
the separation between
the two lowest conduction bands is merely an artifact of poor 
k-point sampling of the Brillouin zone. 
We test whether the EDOS computed using a 32 molecule
supercell with 8 k-points is an accurate approximation of 
the EDOS computed using a 256 molecule supercell with 1 k-point.   
Such calculations would yield identical results for a periodic crystal;
however, for a disordered phase like liquid water this equivalence is destroyed.
Yet, at 300~K the differences are small, and we find that increased 
k-point sampling in smaller supercells
is indeed an accurate approximation of the EDOS of much larger liquid water
systems. This indicates that our converged, k-point sampled EDOS 
of liquid water is accurate
and that the presence of an isolated LUMO of water is spurious.

Our analysis has important consequences for the calculation of spectral
properties (both optical and x-ray) of pure water and materials in
aqueous solvation, since the accuracy of these quantities is dependent on
a reliable description of the unoccupied EDOS. 

The outline of our work is as follows: 
Before presenting our results, we discuss some possible misunderstanding 
of the conduction band in water, which exists in the literature 
(Section~\ref{Sec.IsolatedLUMO} ).
To this end, we build some intuition on the electronic structure of 
liquid water
based on an isolated water dimer and a crystalline phase of ice 
(Sections~\ref{Sec.Dimer} 
and~\ref{Sec.Ice}, respectively).
We efficiently generate a large number of molecular configurations of liquid
water, for use in DFT calculations, using classical molecular dynamics, as
indicated in Section~\ref{Sec.Classical}.
In Sections~\ref{Sec.WaterBands} and~\ref{Sec.DOSkpts} 
we make use of these MD trajectories to
illustrate that, while the $\Gamma$-point
approximation is valid for Brillouin zone integration in
the occupied subspace of Kohn-Sham eigenstates in liquid water, this is
in fact a very poor approximation for the unoccupied states.
To indicate the accuracy of the electronic structure computed with
increased k-point sampling, we compare in Section~\ref{Sec.DOSsize}
with the EDOS of larger supercell calculations.
We outline in Section~\ref{Sec.Consequences} the implications that such
improvements in the description of the electronic structure will have
on calculations which make use of the unoccupied EDOS, 
and in Section~\ref{Sec.Efficiency} we 
show that our approach is quite efficient as a means of increasing the
accuracy of these calculations.
Finally, we give our conclusions in Section~\ref{Sec.Conclusions}.

\section{Previous descriptions of the lower Conduction Band of liquid Water}
\label{Sec.IsolatedLUMO}

\begin{figure}
  \resizebox{\columnwidth}{!}{\includegraphics{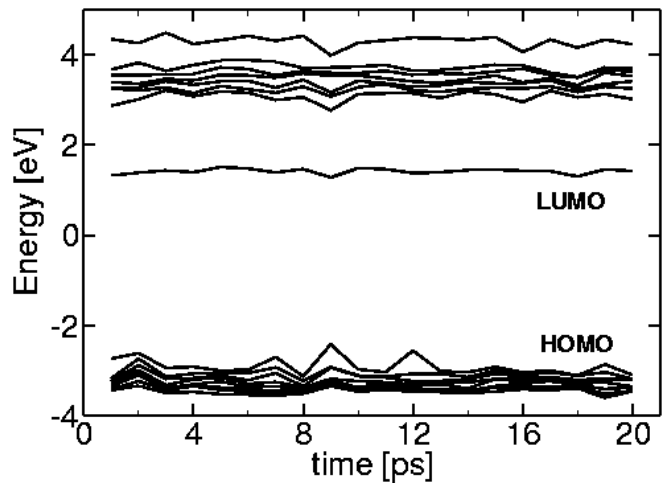}}
  \caption{Occupied and unoccupied Kohn-Sham eigenenergies in
           the neighborhood of the Fermi energy (HOMO),
           computed using DFT/PBE for sequential molecular configurations
           of 32 water molecules taken from a 20~ps Car-Parrinello MD 
           simulation at 300~K, using only the $\Gamma$-point.
           There exists a clear separation in energy between the LUMO and the
           rest of the conduction band.}
  \label{fig.lone_lumo}
\end{figure}

The earliest {\it ab initio} calculations 
on liquid water~\cite{Laasonen_Sprik_1993_jcp} reported the
existence of a delocalized lowest unoccupied molecular orbital (LUMO).
This LUMO 
comprised some molecular $\sigma^*$ character on the oxygens of each water
molecule and a delocalized tail which extended throughout the system in
the intermolecular volume, avoiding hydrogen-bonds. 
In addition, the reported electronic
density of states (EDOS) for a supercell of 32 heavy-water molecules (D$_2$O),
calculated within DFT using the Becke-Perdew gradient corrected exchange 
correlation functional~\cite{Becke_1988_pra,Becke_1992_jcp,Perdew_1986_prb}
reveals an unexpected, isolated state at the bottom of the conduction band.
This is perhaps counterintuitive, as one might expect that a molecular system
would possess
bands of almost degenerate states arising from weak coupling of the
original isolated molecular orbitals; therefore, the conduction band minimum
would be the lowest in energy of a band of states produced by coupling 
of the LUMOs of single water molecules.

Later work~\cite{Boero_Terakura_2001_jcp} displayed the Kohn-Sham
eigenenergies of water, as a function of time during a Car-Parrinello
molecular dynamics~\cite{Car_Parrinello_1985_prl}
simulation, at densities lower than the ambient
density. This showed the isolated peak above the Fermi energy of liquid water
to be caused by a single Kohn-Sham eigenstate, whose separation from the rest
of the conduction band grows with water density. The authors of this work
infer that this ``delocalized LUMO can be regarded as the precursor of the 
solvated electron.'' At these low densities they state that this LUMO has
more delocalized character than those other virtual orbitals higher in
energy, which possess more molecular character.
Subsequent work~\cite{Boero_Parrinello_2003_prl} simulated the occupation
of this LUMO with an
excess electron, performing Car-Parrinello MD at ambient conditions, 
leading to the formation of a
localized region of charge at intervals of approximately 0.1~ps
mediated by the breaking of hydrogen bonds.
The calculated optical absorption from these simulations 
showed good agreement with hydrated excess
electon experiments.~\cite{Jou_Freeman_1979_jpc}

Since then, a large body of work on water and solvated systems has
generated similar time-dependent EDOS, all exhibiting a clear
separation of the LUMO of water from the rest of the conduction band.
~\cite{Tuckerman_Laasonen_1994_jpcm,
       Bernasconi_Sprik_2003_jcp,
       Bernasconi_Blumberger_2004_jcp,
       Prendergast_Grossman_2004_jacs}
For clarity, we report in Fig.~\ref{fig.lone_lumo}
a similar set of Kohn-Sham eigenenergies
as a function of time for a Car-Parrinello MD 
simulation of 32 water molecules at ambient
density ($\rho$=1.0g/cm$^3$) and temperature (300~K).
The Car-Parrinello trajectory was generated using the 
{\tt GP} code~\cite{gp} and
the Kohn-Sham eigenenergies of the unoccupied states for the 
configurations taken from this trajectory
were computed using {\tt ABINIT}.~\cite{abinit}
Both sets of calculations used DFT in the generalized gradient approximation
(GGA) of Perdew, Burke and Ernzerhof (PBE).~\cite{Perdew_Burke_1996_prl}
These were planewave, norm-conserving pseudopotential,~\cite{Hamann_1989_prb} 
electronic structure calculations
performed in a supercell under periodic boundary conditions.
We truncate the planewave basis using a kinetic energy cut-off of 70~Ry
and the oxygen pseudopotential is non-local in the $s$ 
angular momentum channel. 
The Brillouin zone integration is approximated using a single k-point,
${\bf k}={\bf 0}$ (the $\Gamma$-point). This is a standard approximation
for reasonably large supercells of insulating systems and is the same
approach used in previous work, which led to EDOS of water similar
to that of Fig.~\ref{fig.lone_lumo} with an isolated LUMO state.

The presence of such an isolated LUMO is not apparent from the
experimentally measured optical spectrum of 
water,~\cite{Bernas_Ferradini_1997_cp,
Hayashi_Watanabe_2000_pnas} which displays no noticeable peak near the onset
of absorption. This, however, does not discount the presence of such a peak 
in the unoccupied EDOS, which could be diminished in the spectrum by reduced
oscillator strengths associated with optical transitions from the top of
the valence band.
In the following sections we shall examine the electronic structure of
the water dimer (the smallest hydrogen bonded water system) and of
hexagonal ice (a fully saturated hydrogen bonded phase).
From this analysis we shall develop an improved picture of the
unoccupied EDOS of liquid water and show, ultimately, that there is
no isolated peak in the EDOS at the bottom of the conduction band.

\section{The water dimer}
\label{Sec.Dimer}

\begin{figure}
  \resizebox{\columnwidth}{!}{\includegraphics{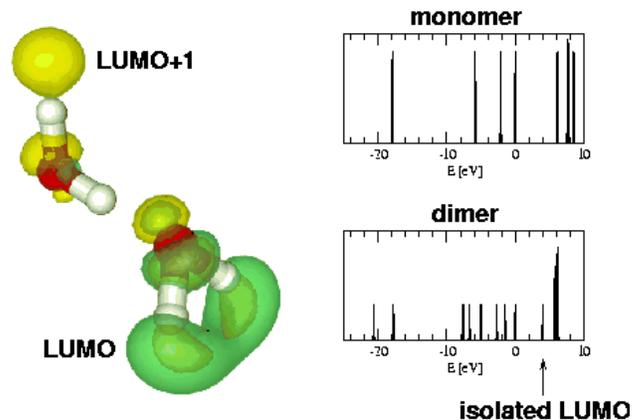}}
  \caption{The water dimer in its ground state geometry computed using
           DFT/PBE. 
           Left: Isosurfaces of the probability densities of the
           first two unoccupied Kohn-Sham eigenstates above the Fermi 
           energy -- LUMO and LUMO+1.
           Water molecules
           are indicated using red (oxygen) and white (hydrogen)
           ball-and-stick models.
           Right: A comparison of the Kohn-Sham eigenenergies
           of the monomer and the dimer
           indicating a separation in energy between LUMO and LUMO+1.
           Note that the energies in both cases have been shifted to align
           the HOMO with zero.}
  \label{fig.dimer}
\end{figure}

Past justifications for the presence
of an isolated LUMO in the u-EDOS of liquid water emphasized comparisons 
with the molecular orbitals of the water dimer. 
We used DFT/PBE to compute the
electronic structure of the water monomer and dimer.
Note that all subsequent DFT calculations outlined in this work
were performed using the {\tt PWSCF} package~\cite{pwscf} and with
a planewave cut-off of 85~Ry,
unless stated otherwise .
Two Kohn-Sham eigenstates of the dimer, the LUMO and
the Kohn-Sham eigenstate just above it in energy (LUMO+1), are indicated
by isosurfaces of their probability densities in Fig.~\ref{fig.dimer}. It is
observed by comparison of the Kohn-Sham eigenenergies 
of the dimer and monomer that the
LUMO is lower in energy than the LUMO+1 as a consequence of being
localized on the dangling hydrogen bonds in this system.
Similarly, in the case of liquid water, the LUMO is 
seen~\cite{Laasonen_Sprik_1993_jcp} to possess a
similar $\sigma^*$ component on the oxygens in the system and also
a significant probability density in those regions where the
hydrogen-bond network is disrupted.

\section{Band structure of ice $\mbox{Ih}$}
\label{Sec.Ice}
 
\begin{figure}
  \resizebox{\columnwidth}{!}{\includegraphics{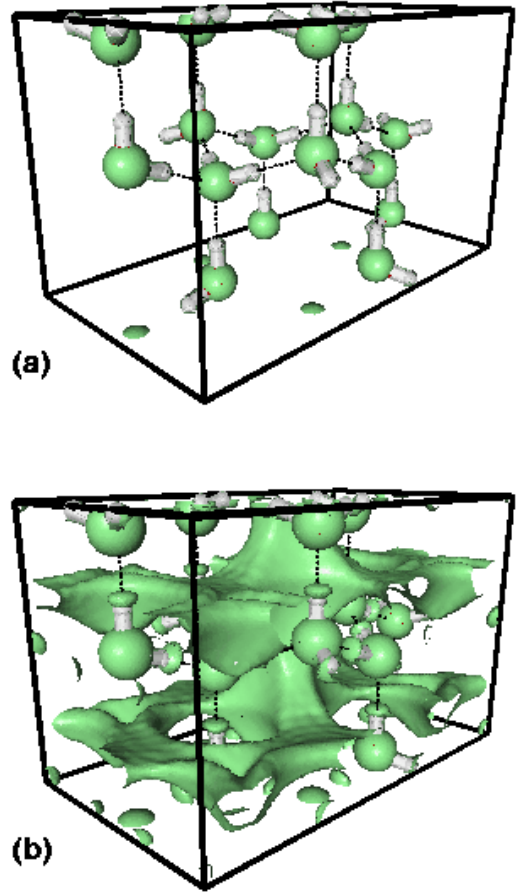}}
  \caption{The ice I$h$ structure with isosurfaces of the 
           probability density (green) of the Kohn-Sham eigenstate just above 
           the Fermi energy at the $\Gamma$-point: 
           (a) the oxygen $\sigma^*$ component; and (b) the delocalized, 
           hydrogen-bond-avoiding component, corresponding, respectively, 
           to $\sim 5$\% and $\sim 30$\% of the integrated density of the 
           state.
           Water molecules
           are indicated using red (oxygen) and white (hydrogen) 
           ball-and-stick models. Hydrogen bonds indicated as dashed lines.}
  \label{fig.ice_lumo}
\end{figure}
 
\begin{figure}
  \resizebox{\columnwidth}{!}{\includegraphics{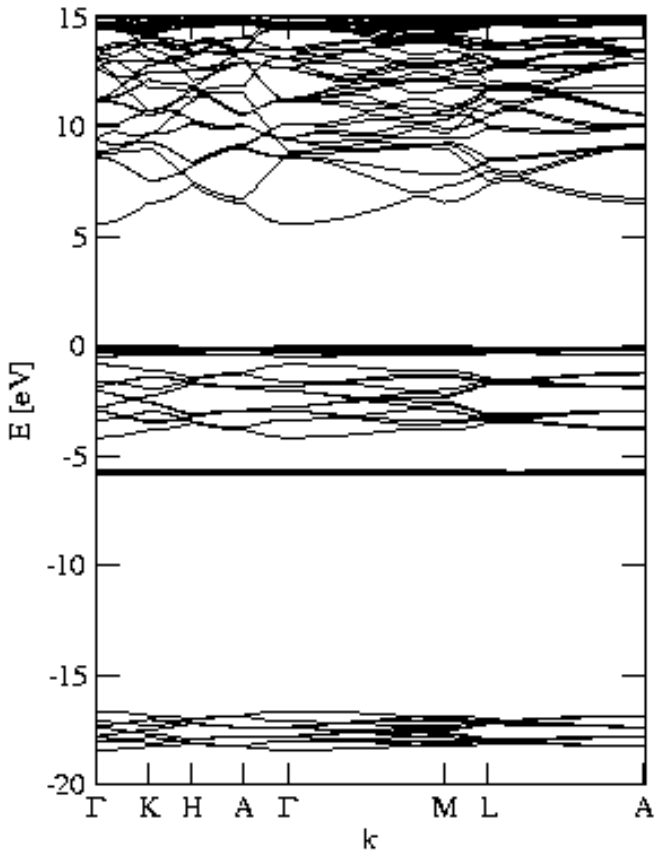}}
  \caption{The band structure of ice I$h$ computed using DFT/PBE
           for the 12 molecule unit cell, with structural parameters
           optimized to reduce the pressure to $\sim 0.1$~MPa.
           The electronic charge density is determined using 8 k-points 
           in the first Brillouin zone.
           (see Fig.~\ref{fig.ice_lumo} and text).}
  \label{fig.icebands}
\end{figure}

Liquid water is perhaps more akin to ice than a gas phase dimer.
In ordered phases of ice, all hydrogen bonds are saturated (i.e., four
hydrogen bonds per water molecule)
and, molecular dynamics simulations of liquid water at ambient
conditions indicate that
the average number of hydrogen bonds per molecule 
is between 3 and 4, depending on the particular definition of
a hydrogen bond.~\cite{Jedlovszky_Brodholt_1998_jcp}
Therefore, before analyzing liquid water in detail, 
we choose to analyze the electronic structure of ice.

Previous work~\cite{Pastori_Parravicini_1973_prb,Xu_Ching_1994_prb} 
on the band structure of cubic ice (I$c$) and the density of states of 
hexagonal ice~\cite{Bai_Su_2003_jcp} (I$h$) gives no indication of
peculiarity at the conduction band minimum.
However, the band structures indicate significant dispersion in the
bands just above the Fermi energy.
Using the hexagonal unit cell of ice I$h$ as proposed by Bernal and
Fowler~\cite{Bernal_Fowler_1933_jcp}, we use DFT/PBE calculations
to generate the structural parameters within the Born-Oppenheimer
approximation at a pressure and temperature of $\sim 0.1$~MPa and 0~K,
respectively.
Proton disorder is not considered while we use the primitive unit cell
containing 12 water molecules with all hydrogen bonds passivated
(Fig.~\ref{fig.ice_lumo}).
This structure has a density of 1.002~g/cm$^3$, corresponding to 
hexagonal lattice parameters of $a=$~7.59~\AA~  and $c=$~7.18~\AA.
The use of the PBE functional is justified by previous work on
ice at various pressures.~\cite{Hamann_1997_prb}
We present the band structure in Fig.~\ref{fig.icebands}. The self-consistent
electronic charge density is computed using 8 k-points in the Brillouin zone
and all eigenenergies in the band structures computed non-self-consistently
using this charge density and its associated Kohn-Sham potential.
Our band structure is similar to that reported by Hahn 
{\it et al.},~\cite{Hahn_Schmidt_2005_prl}
where they employ a cubic cell with some proton disorder, 
and we use the primitive hexagonal cell with no proton disorder.

We note immediately that the large degree of dispersion in the
unoccupied subspace leads to a separation of the lowest conduction 
bands similar to that between the LUMO and
LUMO+1 in previous calculations for water. 
This separation is particularly large at the $\Gamma$-point.
It is clear that considering only the band structure at the $\Gamma$-point
in Fig.~\ref{fig.icebands} would lead to a similar EDOS as that 
outlined for liquid water in Fig.~\ref{fig.lone_lumo}.
Also, the Kohn-Sham eigenstate at the bottom of the conduction band is
delocalized, with $\sigma^*$ character on the oxygen atoms and an avoidance
of hydrogen bonds, similar to the LUMO in previous water calculations.
Furthermore, the separation between the lowest two conduction bands at 
the $\Gamma$-point is 3.0~eV. This large energy is consistent with the trends
indicated by Boero {\it et al.} for the variation of this separation
with density.~\cite{Boero_Terakura_2001_jcp}
The discrepancy between this large separation in ice I$h$ and that of 
$\sim1.5$~eV in ambient liquid water (which has a similar density) is likely 
due to the saturation of all hydrogen bonds in the former, leading
to a more delocalized state and consequently more dispersion at the
$\Gamma$-point.

\section{Classical TIP4P trajectories for liquid water}
\label{Sec.Classical}

\begin{figure}
  \resizebox{\columnwidth}{!}{\includegraphics{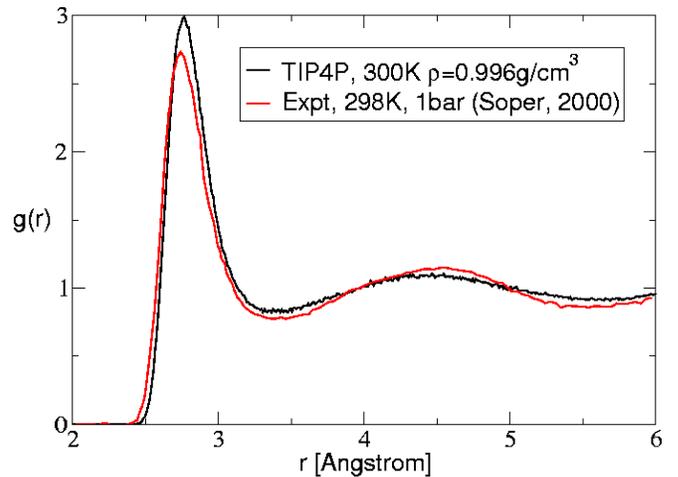}}
  \caption{The oxygen-oxygen radial distribution function of liquid water at
           ambient conditions determined 
           experimentally~\cite{Soper_2000_chemphys} (red) and from
           a molecular dynamics simultion using the TIP4P
           classical potential.}
  \label{fig.tip4p_gor}
\end{figure}

In an attempt to facilitate the efficient reproduction of our results, 
we use configurations of water molecules generated using a classical potential.
This removes the computational expense of generating such configurations using
{\it ab initio} MD, and also provides configurations with structures
in closer agreement with experiment, at least for such measures as the radial
distribution functions and diffusion coefficient. 
Recent, careful, {\it ab initio} DFT/GGA MD simulations 
have been shown to produce more structured radial distribution functions in
comparison with experiment.~\cite{Asthagiri_Pratt_2003_pre,
Grossman_Schwegler_2004_jcp,Schwegler_Grossman_2004_jcp}
We use the TIP4P four-site model~\cite{Jorgensen_Chandrasekhar_1983_jcp}, 
which has been shown recently to have a wide range of transferability 
for water in various condensed phases.~\cite{Sanz_Vega_2004_prl}
We see in Fig.~\ref{fig.tip4p_gor} that TIP4P approximates well the 
experimentally determined~\cite{Soper_2000_chemphys} oxygen-oxygen
radial distribution function of water at ambient conditions.

We used the {\tt gromacs} molecular dynamics package~\cite{gromacs1,gromacs2} to
generate long trajectories for large supercells of water molecules in the
$NVT$ ensemble. We choose the density to be 0.996 g/cm$^3$ and the temperature
to be 300~K. MD simulations for all supercell sizes reported here are begun
using a box of water molecules cut from a large (2048 molecule) equilibrated
sample. The molecules in this supercell are provided with a Boltzmann velocity
distribution consistent with a temperature of 300~K and allowed to equilibrate
for 50~ps. A further 200~ps is evolved for sampling purposes.
All configurations used for DFT calculations are separated by 20~ps.
Previous work indicates that this separation is of the same order of
magnitude as the structural correlation time.
~\cite{Grossman_Schwegler_2004_jcp}

\section{``Band structure'' of liquid water}
\label{Sec.WaterBands}

\begin{figure}
  \resizebox{\columnwidth}{!}{\includegraphics{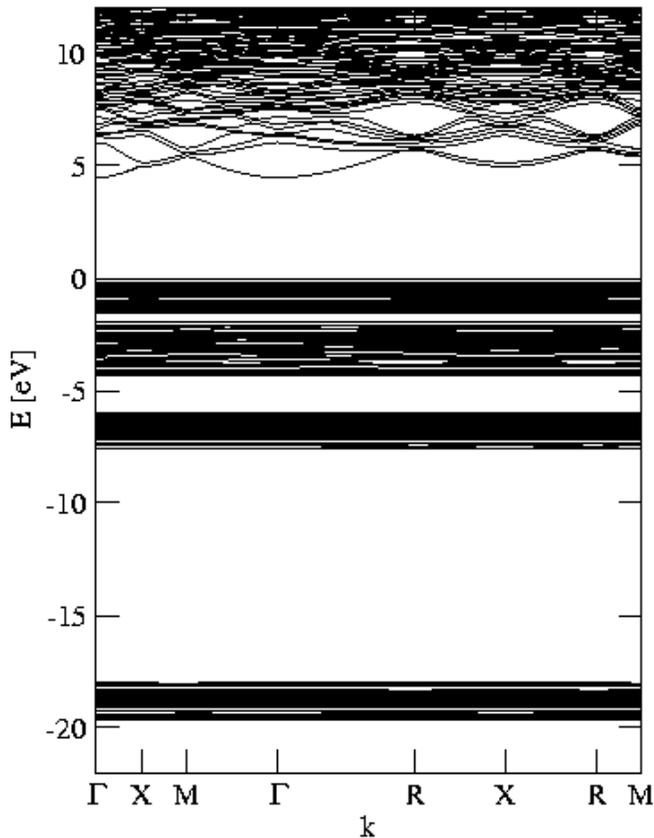}}
  \caption{Computed DFT/PBE ``band structure'' for one representative 
           32 molecule supercell
           of liquid water taken from a TIP4P trajectory.
           The electronic charge density is converged using the
           ${\Gamma}$-point only.}
  \label{fig.water_bands}
\end{figure}

We provide in Fig.~\ref{fig.water_bands} the DFT/PBE calculated
``band structure'' for a representative cubic supercell of 
32 water molecules extracted from the equilibrated section
of a TIP4P classical trajectory.
We recognize that the concept of a band structure has no meaning for
an aperiodic system, however the band structures of the periodic
approximations to the true disordered systems provide information about
the electronic structure of the liquid.
Convergence of the electronic charge density with respect to k-point
sampling in self-consistent calculations for this system
indicated that using just the
$\Gamma$-point is a valid approximation rather than attempting 
a full integration of the Brillouin zone. 
This fact is clearly evidenced by the lack of significant
dispersion in the occupied subspace of this liquid water configuration.
We computed the Kohn-Sham eigenvalues at each k-point in this band structure
using the effective potential derived from the electronic charge density
computed using the $\Gamma$-point approximation.

Examination of the unoccupied bands in this small liquid water supercell
reveals large dispersion, particularly at the $\Gamma$-point.
This indicates that the u-EDOS of this periodic system will be sensitive to the
amount of k-point sampling employed. Clearly, using the EDOS
at the $\Gamma$-point will lead to the familiar separation of states
at the bottom of the conduction band as seen in Fig.~\ref{fig.lone_lumo}. 
However, we can see now that
increasing k-point sampling will add to the EDOS in the region
that has traditionally separated the LUMO and LUMO+1 in
$\Gamma$-point calculations of liquid water using supercells containing
32 water molecules.
We examine this convergence in the next section.

\section{Convergence of the EDOS with k-point sampling}
\label{Sec.DOSkpts}

\begin{figure*}[!]
  \resizebox{\textwidth}{!}{\includegraphics{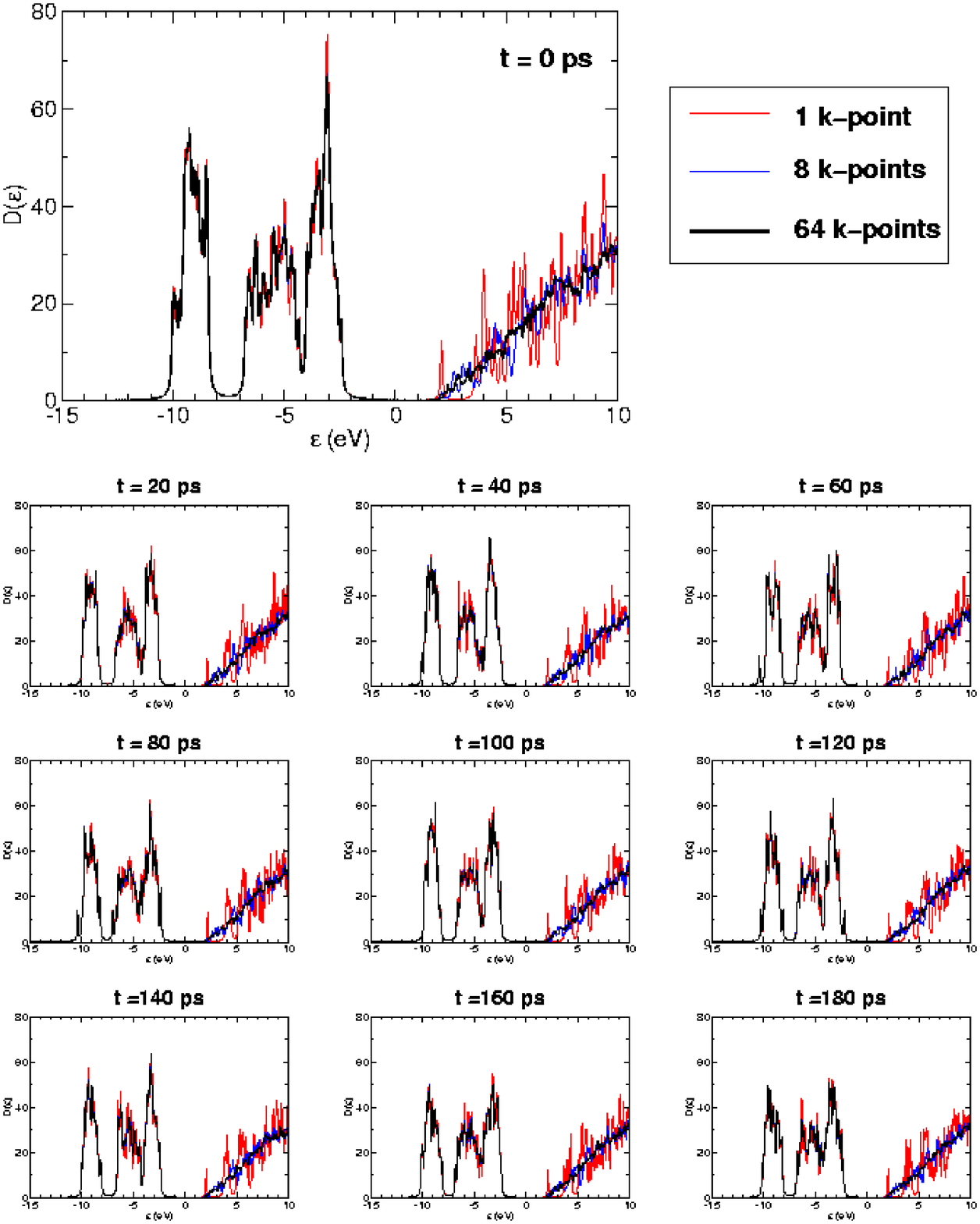}}
  \caption{Convergence of the EDOS, computed within DFT/PBE, 
           with respect to k-point sampling
           for 10 uncorrelated configurations of 32 water molecules
           sampled every 20ps from a 200~ps TIP4P, $NVT$ molecular dynamics
           trajectory at T=300~K, $\rho=0.996\mbox{g/cm}^3$.
           The EDOS are broadened using Lorentzians with a full width at half
           maximum of 0.05~eV.}
  \label{fig.dos_kpts}
\end{figure*}

We take 10 uncorrelated molecular configurations of 32 water molecules
of liquid water, at ambient conditions, at intervals of 20~ps from
a TIP4P trajectory. For each of these we perform DFT/PBE electronic structure
calculations, increasing the k-point sampling of the first Brillouin zone.
The indicated number of k-points refers to the number of points
in a uniform grid about ${\bf k} = {\bf 0}$. For example, 8 k-points
implies a $2\times2\times2$ grid. However, using the symmetry of
our cubic supercells we reduce the actual number of k-points used
in the calculation, reweighting each appropriately in the sum which 
approximates a complete Brillouin zone integration. In each of the
configurations examined we find that 64 k-points is sufficient to
converge the EDOS to the accuracy necessary for this demonstration.
Note that more rapid convergence can sometimes be achieved by generating
k-point grids centred about a k-point other than $\Gamma$. We did
not test such grids in this work.

The results of these convergence tests are displayed in 
Fig.~\ref{fig.dos_kpts}. Close examination of the enlarged EDOS
at t~=~0~ps demonstrates that the occupied EDOS (below zero in energy)
remains essentially unchanged with k-point sampling. The only
noticeable effect is some reduction in the sharpness of features when using
the $\Gamma$-point with small numerical broadening (0.05~eV in this case). 
However, we see clearly that the unoccupied EDOS
is greatly modified as the k-point density is increased.
The gap that exists between LUMO and LUMO+1 (from 2~eV to 3.5~eV)
under the $\Gamma$-point approximation is filled completely at
higher k-point densities. Furthermore, the qualitative form of 
the EDOS beyond the conduction band minimum is completley different.
Comparison with the other uncorrelated molecular configurations
reveals the same behavior. In fact, as a function of time, the
converged EDOS shows more variation below the Fermi energy 
than it does in the unoccupied subspace. These variations are likely
due to particular relative orientations of water molecules or
making and breaking of hydrogen bonds, but will not be investigated
here.

\section{Convergence of the EDOS with system size}
\label{Sec.DOSsize}

\begin{figure*}[!]
  \resizebox{\textwidth}{!}{\includegraphics{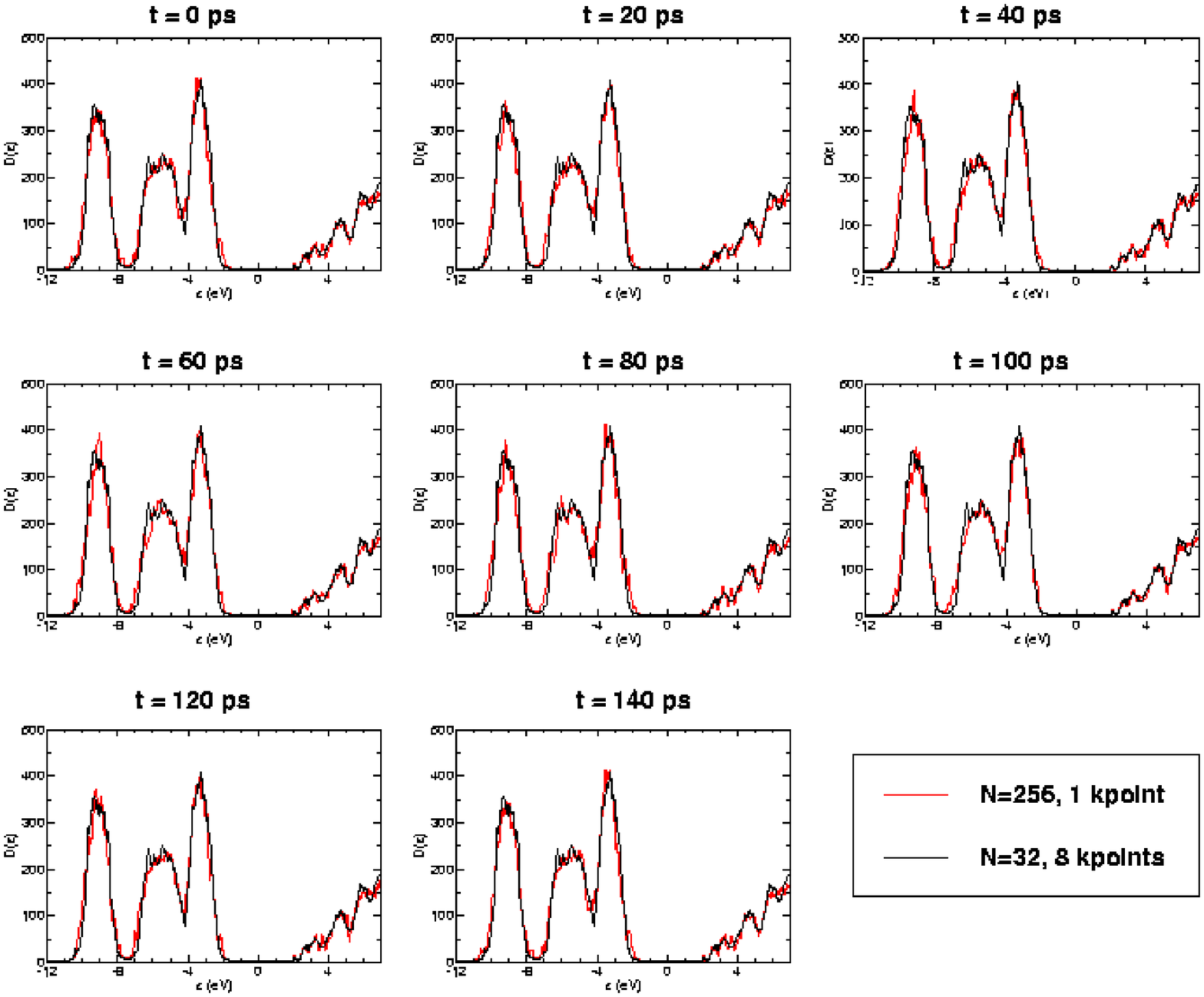}}
  \caption{Comparison of EDOS for liquid water, computed within DFT/PBE,
           from 256 molecule supercells,
           sampled every 20ps from a 140~ps TIP4P, $NVT$ molecular dynamics
           trajectory at T=300~K,
           $\rho=0.996\mbox{g/cm}^3$, with a time averaged EDOS from an
           uncorrelated TIP4P trajectory using a 32 molecule supercell, 
           computed using 8 k-points in the
           first Brillouin zone.
           The EDOS are broadened using Lorentzians with a full width at half
           maximum of 0.05~eV.}
  \label{fig.dos_size}
\end{figure*}

For an electronic structure calculation of a periodic system, 
accurate Brillouin zone integration may be achieved
either by (i) increasing the k-point sampling of the Brillouin zone 
of the primitive unit cell, or (ii) by using larger supercells comprising
repeated unit cells and a minimal k-point sampling.
The latter approach is clearly more expensive and where possible we would
prefer the former option.
However, for disordered systems, the equivalence of these two approaches
no longer holds, since the system is no longer periodic.
Therefore, it is, in principle, more accurate within calculations performed
under periodic boundary conditions to approach the limit of the
bulk, disordered phase by increasing the supercell size and using the
$\Gamma$-point approximation. This is also a requirement for the
analysis of long-range molecular structure in the disordered phase.

However, experience from tight-binding calculations of a variety of systems,
indicates that the electronic structure of homogeneous systems
is, perhaps, influenced most by short-range atomic structure, up to
second-nearest-neighbor interactions. Therefore, we used small,
32 molecule supercells and increased k-point sampling to approximate
the electronic structure of larger disordered structures.
Such comparisons are consistent when the k-point density is 
the same.~\cite{Makov_Payne_1995_prb,Makov_Shah_1996_prb}
This guarantees that the electronic degrees of freedom encompass the
same volume in real-space, the difference being that the system with
larger k-point density possesses more structural order, and so is a
poorer approximation to the disorderd liquid.

In Fig.~\ref{fig.dos_size} we present the EDOS computed using DFT/PBE
for large, 256 water molecule supercells. These eight uncorrelated
configurations
of liquid water are taken from another independent TIP4P trajectory,
prepared as outlined in Section~\ref{Sec.Classical}.
These EDOS are computed using the $\Gamma$-point approximation and
compared with an EDOS representative of a 32 molecule liquid water
supercell using 8 k-points. This EDOS for the smaller system is
actually the average of the 8-kpoint EDOS over the 180~ps
presented in Fig.~\ref{fig.dos_kpts}.
We see that the essential features of the EDOS of the larger system
are well reproduced by the smaller system with an equivalent k-point density.
In fact the agreement is remarkable considering that there is
no correlation between these structures and that the data spans
140~ps. 

The implications of this agreement are that smaller systems
with increased k-point density can accurately represent the electronic
structure of larger liquid water supercells. Furthermore, since we
require at least 64 k-points to converge the EDOS of the 32 molecule
configurations (Fig.~\ref{fig.dos_kpts}), 
it is clear that the 256 molecule, $\Gamma$-point
calculations are still not converged. The computational
expense of generating DFT/GGA electronic structures for such large
systems prohibits the comparison with supercells containing
2048 water molecules.
Fortunately, it seems that such exhorbitant calculations are
unnecessary for accurate prediction of (at least) the DFT electronic stucture
of liquid water.

\begin{figure}
  \resizebox{\columnwidth}{!}{\includegraphics{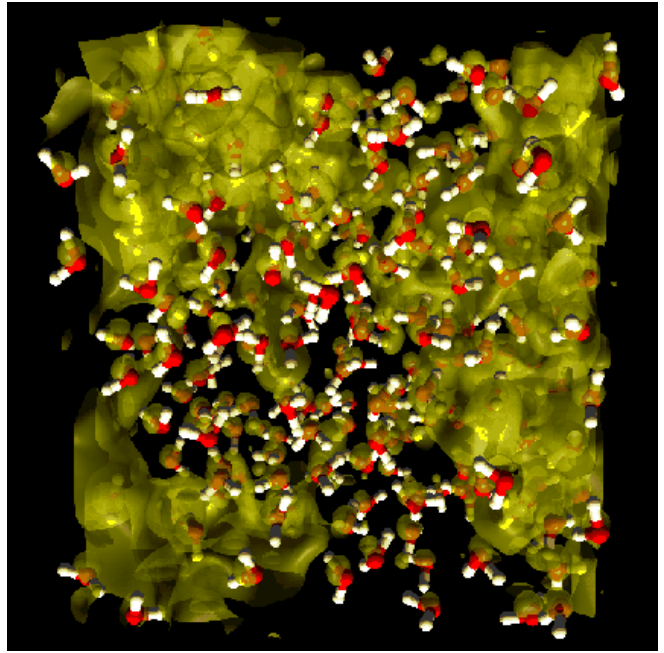}}
  \caption{An isosurface (gold) of the probability density of the 
           Kohn-Sham eigenstate at the conduction band minimum
           of a 256 molecule supercell of liquid water
           extracted from a TIP4P trajectory. Water molecules
           indicated as red (oxygen) and white (hydrogen) ball-and-stick
           models.}
  \label{fig.256lumo}
\end{figure}

We note also that the qualitative features of the Kohn-Sham eigenstate
at the conduction band minimum (Fig.~\ref{fig.256lumo}) of these
very large, 256 molecule supercells
are the same as those of the LUMO computed in the smaller $\Gamma$-point
calculations and in previous work. 

\section{Consequences for solvation and spectroscopy}
\label{Sec.Consequences}

\begin{figure}
  \resizebox{\columnwidth}{!}{\includegraphics{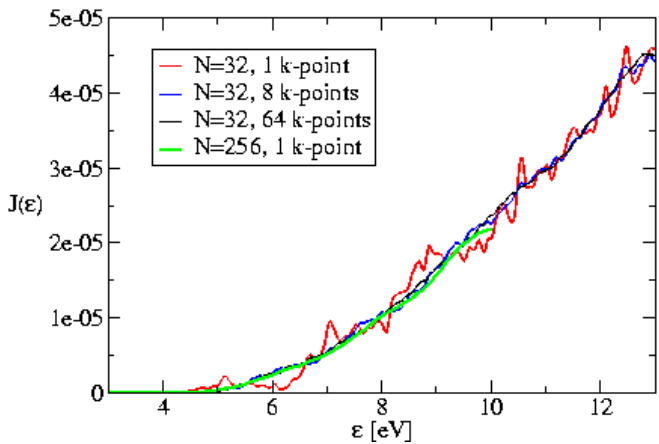}}
  \caption{Convergence of the joint density of states (JDOS) of liquid water
           (arbitrary units):
           for a 32 molecule supercell with 1 k-point (red); 8 k-points (blue);
           64 k-points (black); and for a 256 molecule supercell with
           1 k-point (green) with data only up to 10~eV excitations due to
           a limited number of unoccupied states in this calculation.
           A Gaussian broadening of 0.05~eV has been employed 
           to expand the JDOS associated
           with each specific transition.}
  \label{fig.jdos}
\end{figure}

At least for the electronic ground state of molecular systems, 
our calculations show
that $\Gamma$-point sampling is sufficient for convergence of the DFT
electronic structure. However, care should be taken to verify that
the dispersion in such systems is in fact minimal in order to justify this
approximation. In particular, simulations involving phase transitions from
disordered to ordered phases or from dilute to concentrated phases may 
exhibit different degrees of
dispersion in the occupied EDOS, and the minimum k-point sampling required
for the more dispersive phase should be adopted.

The large degree of dispersion in the lowest conduction band
of liquid water (Fig.~\ref{fig.water_bands}) may have consequences
in the simulation of a hydrated, excess electron.
~\cite{Boero_Parrinello_2003_prl}
In the limit of large system size, there should be a continuum of
states at the conduction band minimum, allowing for the possibility of
``intraband'' transitions mediated by finite temperature.
It is not clear what impact such transitions may have on the dynamics
of such a system and the time scale for localization of the solvated electron.

Of course, all of our work is limited by the accuracy of DFT, and, for excited
electronic states, DFT has well-recognized limitations. 
This has been explicitly demonstrated for ice I$h$ in a recent
publication.~\cite{Hahn_Schmidt_2005_prl} However, 
many-body approaches, such as GW,~\cite{Hybertsen_Louie_1986_prb} 
which allow for improvements in the description
of the spectrum of excitations, rely on the use of Kohn-Sham eigenstates as
a starting point. Therefore, issues concerning the supercell size and
k-point sampling are also relevant for these calculations. In particular,
if the size of supercell can be reduced in favor of increased k-point
sampling, this will be extremely advantageous for these computationally
expensive methodologies.

DFT estimates of the optical absorption spectrum of liquid water will
also be sensitive to the system size. We illustrate this in Fig.~\ref{fig.jdos},
where we compare the joint density of states (JDOS) of water for 
various k-point sampling schemes and system sizes.
The JDOS is a first estimate of the optical absorption spectrum, ignoring
the role of symmetry in electronic transitions from valence to conduction band.
We compute the JDOS by considering only interband transitions from the
valence to the conduction band.
We notice that the JDOS of liquid water is less sensitive to k-point sampling
than the underlying EDOS (at least for the broadening scheme we have adopted
in Fig.~\ref{fig.jdos}). 
This is to be expected given that the JDOS is
a convolution of the valence and conduction band EDOS.
Furthermore, the same transferability to larger systems is apparent when we
compare the JDOS of a 256 molecule liquid water supercell with that
of an uncorrelated, 32 molecule supercell using 8 k-points.
We note that both of these JDOS are qualitatively different from that
computed using the smaller supercell within the $\Gamma$-point approximation.
The existence of a peak at the absorption onset is one of the expected
consequences of poor Brillouin zone sampling.

DFT investigations into the optical properties of molecules and ions
in aqueous solution also require particular attention with respect to
accurate representations of the electronic structure of water.
Let us ignore, for the moment, the strong possibility of different 
systematic band gap errors for the solute and solvent within DFT,
as has been demonstrated using hybrid exchange correlation 
functionals.~\cite{Bernasconi_Sprik_2004_cpl}
If there exists a hybridization of solute and solvent states, which would modify
the optical properties, this may be inaccurately estimated using DFT
under the $\Gamma$-point approximation, particularly if the relevant,
optically active
solute state mixes with the bottom of the conduction band of water.
~\cite{Prendergast_Grossman_2004_jacs,
       Bernasconi_Sprik_2003_jcp,
       Bernasconi_Sprik_2004_cpl,
       Bernasconi_Blumberger_2004_jcp}
Under the $\Gamma$-point approximation this state can hybridize with  
only one water state,
whereas our analysis indicates the presence of a continuum of states
in this energy range.

\begin{figure}[t]
  \resizebox{\columnwidth}{!}{\includegraphics{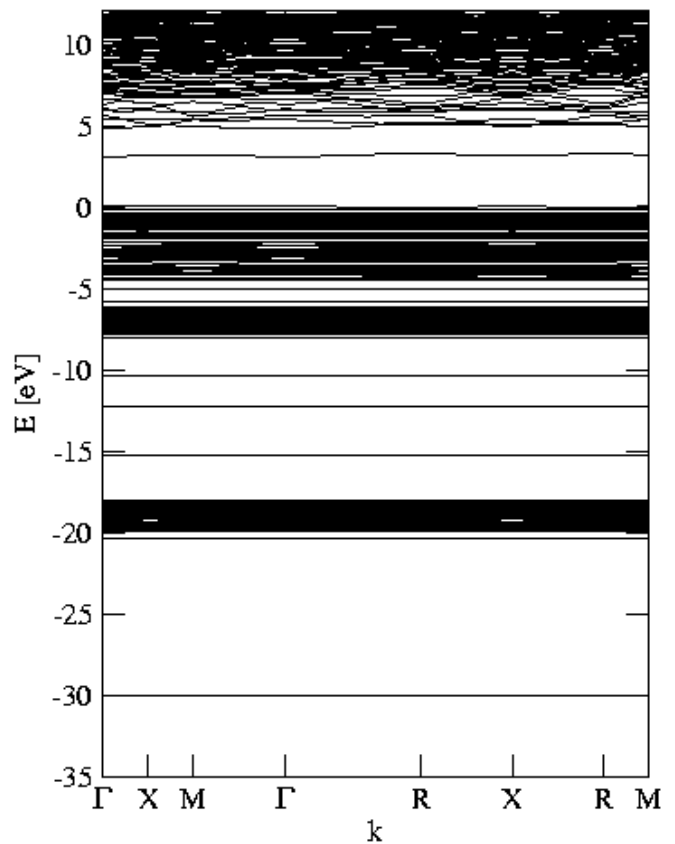}}
  \caption{The ``band structure'' of a 32 molecule liquid water supercell
           in the presence of a core (x-ray) excitation from the $1s$ 
           orbital of one particular oxygen atom modelled using a modified
           pseudopotential (see text).}
  \label{fig.xray}
\end{figure}

Finally, we demonstate (Fig.~\ref{fig.xray}) the degree of dispersion 
in the Kohn-Sham eigenstates
of a 32 molecule supercell of liquid water in the presence of an x-ray
excitation.  
We model this excitation using a modified pseudopotential which includes
a core hole in the $1s$ level for one particular oxygen atom in the system.
The system is effectively ionized and the impact of this perturbation is
apparent when the band structure is compared with the ground state
(Fig.~\ref{fig.water_bands}).
Localized states are realized, shifted from their respective bands.
The conduction band now exhibits states of varying localization, i.e.
a subset exhibiting minimal dispersion and the remaining states retaining
the typical dispersion associated with the ground state conduction band. 
Increased k-point sampling beyond the $\Gamma$-point is required
to accurately describe all of the conduction band which will be
incorporated in spectral calculations for this system.

\section{Computational efficiency}
\label{Sec.Efficiency}

The calculations presented here provide useful guidelines, in general,
for the estimation of the
electronic properties of disordered molecular systems in the condensed phase.
While these facts may be well known, we feel that it is useful to reiterate
them here as they should be applied to liquid water.

Firstly, it is important to know, for a given supercell size, what amount
of k-point sampling is sufficient to describe the occupied subspace of
Kohn-Sham eigenstates. This can be easily checked by analysis of the
band structure of a representative molecular configuration, or, more
consistently, by checking the convergence of the EDOS with respect to
k-point sampling. Note that special care is required for the case of possible
phase transitions, in small supercells, to states where dispersion is large.

Secondly, as we have shown in this work, knowledge of the degree of
dispersion in the occupied subspace is not necessarily transferable to
the unoccupied subspace. It may be that a more dense k-point sampling is
required for the unoccupied bands. However, given that these states
do not influence the electronic charge density of the system, they may
be generated non-self-consistently. One may quickly converge the electronic
charge density of the system by considering only the occupied states
and use a sparse k-point sampling, e.g., just the $\Gamma$-point for a
32 molecule supercell of water.
This charge density may be used to generate the common effective potential in
a large set of Kohn-Sham equations which are only coupled if they correspond
to the same k-point in the Brillouin zone. These equations may be solved
using just one matrix diagonalization each, a process that is trivially
parallelizable with perfect linear scaling with respect to the number of
k-points.

Thirdly, we have shown that the EDOS of a relatively small supercell of
water molecules computed with k-points is a
very good approximation to the EDOS of a larger system,
with an equivalent k-point sampling density. In the case of water, this is
a huge saving in computational cost, since increasing the system size
eightfold introduces an increase in computational time by a factor of 512 
for a typical planewave pseudopotential calculation, which scales at worst
as $O(N^3)$, where $N$ is the number of electrons in the system.
On the other hand, using 8 k-points for the original supercell introduces
only an eightfold increase in computational cost, as we have shown.
Furthermore, we have also shown that such large calculations: 256 molecules
with $\Gamma$-point sampling; are still not converged in the EDOS. So,
if we wish to approach a converged result, increased k-point sampling
may be the only resort. This necessity does not
impose a limit on accuracy in electronic structure calculations for 
liquid water.

\section{Conclusions}
\label{Sec.Conclusions}

In this work, we have provided a clear insight into the electronic
structure of liquid water within the context of density functional theory --
in particular using the PBE, gradient corrected, exchange correlation 
functional.
We have shown that an accurate representation of the electronic structure
is provided by relatively small, 32 molecule supercells of the liquid.
By inspection of the convergence of the electronic density of states (EDOS)
with respect to k-point sampling of the first Brillouin zone, we have
verified that the $\Gamma$-point approximation is adequate for accurate
estimation of the occupied EDOS and consequently all ground state electronic
properties. However, we find that the $\Gamma$-point approximation is
inadequate in providing an accurate description of the unoccupied EDOS
of liquid water, which required a $4 \times 4 \times 4$ k-point mesh
for convergence. Comparison with larger, uncorrelated, 256 molecule 
supercells of liquid water indicates that dense k-point sampling
for small supercells provides an accurate estimation of the electronic structure
of larger supercells which more closely approximate the structural
disorder of the liquid.
This reveals the possibility of marked savings
in computational cost when examining the electronic structure of
molecular liquids such as water and, in particular, when attempting to
use DFT to estimate the spectroscopic properties of such systems.
Work is in progress to calculate both optical and x-ray absorption spectra
for liquid water based on the analysis presented here.

\begin{acknowledgments}

We wish to acknowledge T. Ogitsu, G. Cicero, F. Gygi, and A. J. Williamson 
for useful discussions. 
We thank N. Marzari for his suggestions regarding {\tt PWSCF}.
This work was performed under the auspices of the U.S. Department of Energy
at the University of California/Lawrence Livermore National Laboratory under
Contract No. W-7405-Eng-48.

\end{acknowledgments}

\end{document}